# The magnetic properties of the Garnet and glass forms of $Mn_3Al_2Si_3O_{12}$

G.C. Lau[1], T. Klimczuk[2,3,*], F. Ronning[2], T.M. McQueen[1], and R.J. Cava[1]

[1]Department of Chemistry, Princeton University Princeton NJ 08544

[2]Los Alamos National Laboratory, Los Alamos, NM 87545, USA

[3] Faculty of Applied Physics and Mathematics, Gdansk University of Technology, Narutowicza 11/12, 80-952 Gdansk, Poland

**Abstract**

The magnetic susceptibilities and specific heats of the crystalline garnet and glass forms of $Mn_3Al_2Si_3O_{12}$ are reported. This allows a direct comparison of the degree of magnetic frustration of the triangle-based garnet lattice and the structurally disordered solid at the same composition for isotropic spin 5/2 $Mn^{2+}$ ($3d^5$). The results show that the glass phase shows more pronounced signs of magnetic frustration than the crystalline phase. Through comparison of the specific heats of $Ca_3Al_2Si_3O_{12}$ (grossular) and $Mn_3Al_2Si_3O_{12}$ (spessartine) garnets, information is provided concerning the anomalous extra specific heat in the latter material.

*current address: European Commission, Joint Research Centre, Institute for Transuranium Elements, Postfach 2340, Karlsruhe, D-76125 Germany

**Introduction**

Geometric magnetic frustration in triangular lattices is of substantial current interest. In addition to geometry, disorder on the magnetic lattice is also an important factor in determining the presence or absence of long range periodic magnetic ordering (1-7). Many of the geometrically frustrated crystalline magnets also include some degree of structural disorder, complicating the interpretation of their observed properties. However, there are no reports that present a comparison of the magnetic properties of a structural glass and an ordered crystalline solid with a frustrating crystal structure at the same composition. Here we compare the magnetic properties of the crystalline garnet form of $Mn_3Al_2Si_3O_{12}$ (spessartine) and a structural glass of the same composition. The garnet lattice, made of interpenetrating rings of corner-sharing magnetic triangles, is geometrically frustrating (1,8). Crystalline $Mn_3Al_2Si_3O_{12}$ has been of interest in geochemistry as one end-member of the grossular-spessartine garnet $Ca_{3-x}Mn_xAl_2Si_3O_{12}$ solid solution (see e.g. 9,10), and the glass form of $Mn_3Al_2Si_3O_{12}$ has been studied as a canonical example of an insulating spin glass (see e.g. 11-13). Through the comparison of magnetic susceptibility and specific heat measurements, here we show that for this case the structural glass exhibits signs of greater magnetic frustration than the crystalline phase. Comparison of the specific heats of crystalline $Ca_3Al_2Si_3O_{12}$ and $Mn_3Al_2Si_3O_{12}$ shows that the previously observed anomalous extra contribution to the specific heat in the Mn compound is consistent with either an order-disorder transition of the $Mn^{2+}$ ions (14-16), in our analysis occurring over a temperature range of 30 – 300 K, or the shift of one of the phonon modes of the Mn in spessartine to approximately one third the energy seen for the Ca in grossular.

**Experimental**

$Mn_3Al_2Si_3O_{12}$ was synthesized from a stoichiometric mixture of powders of $MnO_2$, $Al_2O_3$, and $SiO_2$ (Alfa Inorganics, 99.9+% purity) heated in a high density alumina crucible under $N_2$ flow. The mixture was preheated at 900 ºC for several hours and then heated to between 1100-1200 ºC and annealed for 4 hrs. This melt was cooled at about 400 ºC/hr to yield the glass phase (11). Brown crystals of spessartine garnet were obtained by cooling a melt as above, but at 5 ºC/hr, down to 900 ºC, and then furnace cooling (17). The nonmagnetic glass and crystalline equivalents of these materials, employed for subtraction of the lattice part of the specific heat at low temperatures, were synthesized at the grossular composition, $Ca_3Al_2Si_3O_{12}$. $CaCO_3$ was employed as the CaO source. A stoichiometric mixture was first heated in a platinum crucible in air at 1200 ºC. To obtain the glass form, the material was melted under Ar and then cooled quickly to room temperature. The crystalline form was obtained by heating the glass at 1300 ºC for 30 minutes at a pressure of 4 GPa in a boron nitride crucible and cooled to room temperature in approximately 2 hours, under the applied pressure. The purities of the crystalline products and structural glasses were verified by powder X-ray diffraction (Bruker D8 Focus, Cu Kα radiation, graphite diffracted beam monochromator). The specific heat and magnetic susceptibilities were determined using a Quantum Design Physical Property Measurement System (PPMS). An applied field of 1 Tesla was employed in the susceptibility measurements. M vs. H curves were linear up to this field for all temperatures and materials measured.

**Results**

The temperature dependent magnetic susceptibilities, M/H, are presented in Fig. 1 for crystalline and glass forms of $Mn_3Al_2Si_3O_{12}$. The high temperature data follow the Curie-Weiss law, $\chi = C/(T-\theta_{CW})$ in both cases, where $\chi$ is the susceptibility, T is temperature, C is the Curie constant and $\theta_{CW}$ is the Weiss theta. Fits to the data in the temperature range 125-275 K yield $\theta_{CW}$ = -20.8 K and -87.1 K for the crystal and glass, respectively. Similar effective moments are obtained, 5.72 and 5.56 $\mu_B$ respectively, which compare favorably to the spin only value of $3d^5$ $Mn^{2+}$, 5.9 $\mu_B$ (18). The higher negative $\theta_{CW}$ for the structural glass form is consistent with superexchange mediated interactions between $Mn^{2+}$ and the presence of shorter Mn-Mn separations in the glass, 2.8 Å, compared to those in crystalline Mn garnet, 3.5 Å (12). The structural glass shows no long range magnetic ordering at low temperatures (inset Fig. 1) and at these applied fields shows deviation from Curie Weiss behavior near 80 K, a temperature comparable to $\theta_{CW}$. The crystalline phase shows a cusp in susceptibility at 7 K. This is due to the long range ordering of the $Mn^{2+}$ moments, which are reported to assume a scheme in which the corner sharing triangles display opposing magnetic chirality to yield a long range ordered antiferromagnetic state (19-22). Suppression of the magnetic ordering temperature to 1/3 of the Curie Weiss temperature indicates the presence of only a small degree of magnetic frustration in this system ($\theta_{CW}/T_N$ = 3) in spite of the triangle-based magnetic lattice and the isotropic $Mn^{2+}$ spin of 5/2.

The specific heats of the magnetic and nonmagnetic crystalline and glass garnets are presented over a wide range of temperature in Fig. 2. A sharp peak in C/T is seen for crystalline $Mn_3Al_2Si_3O_{12}$, whereas the $Mn_3Al_2Si_3O_{12}$ glass shows higher C/T over a wider range of temperature but no sharp magnetic ordering peak. The non-magnetic analogs show lower C/T at low temperatures but the heat capacities become quite similar at temperatures above about 200 K. The heat capacities of crystalline and glass $Ca_3Al_2Si_3O_{12}$ provide a baseline for consideration of the contribution of the magnetic constituents to the specific heat in the Mn analogs; they are employed to provide the subtraction of the lattice parts of the specific heats.

The difference between the specific heats for the magnetic and nonmagnetic materials for the glass phases is presented in the inset to Fig. 2. The glass phase displays a continuous release of magnetic entropy over a wide temperature range, as expected. A simple subtraction of the Ca glass specific heat from the Mn glass specific heat yields the temperature dependence of the magnetic entropy (inset Fig. 2). The entropy loss is seen to begin at about 100 K, comparable to the temperature where the magnetic data deviates from the Curie-Weiss law. The integrated magnetic entropy loss in the glass is approximately Rln5. Somewhat less than Rln6 (the expected Rln(2S+1) value, where S=5/2 is the spin for $Mn^{2+}$) this reflects the presence of remnant magnetic disorder in the $Mn_3Al_2Si_3O_{12}$ spin glass at low temperatures.

Integration of the total entropy difference between the Mn and Ca crystalline garnets (i.e. C/T (Mn, spessartine) − C/T (Ca, grossular)), yields a value that is substantially larger than is possible for the Rln6 magnetic entropy for magnetic ordering of $Mn^{2+}$. This indicates that there is an additional contribution to the entropy in crystalline $Mn_3Al_2Si_3O_{12}$. This excess entropy is found under a broad peak near 90 K (Fig. 3 inset.) The existence of excess entropy in spessartine is a well-known issue in geochemistry and the observed heat capacity has been fit to various models (10,14-16). The current data, by directly characterizing the difference in low temperature thermodynamic behavior between $Mn_3Al_2Si_3O_{12}$ and $Ca_3Al_2Si_3O_{12}$, provides further insight into this issue.

To explain this broad high temperature peak in the entropy difference, reference to the low temperature crystal structures of $Mn_3Al_2Si_3O_{12}$ and $Ca_3Al_2Si_3O_{12}$ (9,10) is of interest. The structure determinations at 100 K indicate that for both compounds, the large cage sites, in which dodecahedrally coordinated $Mn^{2+}$ and $Ca^{2+}$ ions are found, show disordered distributions of ions among different sites displaced from the cage centers. This displacement is characteristic of what is frequently observed for ions in host frameworks that are positioned in cavities that are too large. The structure determinations cannot specify whether such disorder is static or dynamic; in other words whether an ion is at rest in one of several low energy positions within a cage, and others are at rest in other sites in different cages, or whether the disorder is dynamic, i.e. whether an ion is actively moving ("rattling") among the various low energy sites within a cage.

The temperature dependence of the integrated excess entropy of the crystalline form of $Mn_3Al_2Si_3O_{12}$, obtained from subtraction of C/T for the non-magnetic Ca analog, is shown in Fig. 3. For spessartine, the entropy change occurs in two steps. The low temperature change is that of the magnetic ordering transition at 7 K. The entropy sum for this part is within experimental uncertainty of Rln6, the ideal expected for the full ordering of spin 5/2 moments, in agreement with previous studies (15). The second part, the entropy under the broad peak near 90 K, integrates to approximately Rln4. If this is due to the presence of a freezing transition of the Mn into off center positions in its cage (i.e. a disorder to order transition (14,16)) occurring over a wide range of temperatures, about 300 – 30 K from our data, and assuming that the Ca ions are frozen in their positions at high temperatures, then the integrated entropy is expected to be Rln$n$, where $n$ is the number of Mn positions within the cage. To obtain the observed value of Rln4, the Mn ions would have to displace from cage center 24c sites in the garnet space group to the general positions 96g (16). This appears to be inconsistent with crystallographic studies suggesting that the Mn occupies two positions within the cage (9,10), but detailed structural study at temperatures below 30 K is needed to define the Mn positions more precisely.

An alternative explanation for excess specific heat in the spessartine Mn garnet is that many phonon frequencies remain the same in $Mn_3Al_2Si_3O_{12}$ and $Ca_3Al_2Si_3O_{12}$, however one, associated with the Mn and Ca atoms, changes energy. The specific heat due to that mode, modeled as an Einstein mode, is a continuously increasing function of temperature. It is shown as the green dotted curve in the inset to Fig. 3 for the Mn compound (an Einstein mode is shown, but an equally good description is provided by a Debye mode) (23,24). The same phonon is present in the Ca analog, but occurs at a higher energy. The specific heat for that mode in the Ca sample is shown by the brown dashed curve in the inset. The difference between these curves results in a peak in $C_{Mn} - C_{Ca}$, shown by the black solid curve. By fitting the high temperature peak to this scenario, the Einstein temperatures of the modes are found to be 148 K and 582 K for the Mn and Ca materials respectively, (or alternatively, for Debye modes, the Debye temperatures are 232 K and 680 K). The agreement for the Einstein mode case is shown in the inset to Fig. 3, but both give an excellent agreement with observations. The approximately three-fold decrease in mode frequency is significantly greater than that expected simply due to the differences in masses for Mn and Ca for an oscillator with a fixed spring constant. Thus the spring constant in this scenario would be lowered due to the looser fit of $Mn^{2+}$ in the cage, compared to $Ca^{2+}$. Given that the crystal structure determinations show a substantially more extended atomic density around the Mn sites than around the Ca sites at low temperatures (10), this is a plausible explanation.

The effect of applying a magnetic field on the magnetic transitions in $Mn_3Al_2Si_3O_{12}$ garnet and glass is shown in Fig. 4. The data show that the ordering transition in the Mn garnet phase is affected only slightly in an applied field, with a change at most of a few tenths of a degree in a field of 8 Tesla. Thus the magnetic order in the Mn garnet crystalline phase, in spite of its complexity (19-22), is strongly locked in. For the glass, on the other hand, the entropy loss is shifted to higher temperature with applied field due to the fact that the degeneracy of states at low temperatures is being lifted by the magnetic field, as is typically observed in spin glasses. The residual disorder seen at 0 T should eventually be recovered with increasing field, which is the trend in the current data, but further work would be needed to clarify this behavior.

Finally, direct comparison of the magnetic behavior of the glass and crystal Manganese garnets can be facilitated by suitably normalizing the susceptibility data. Starting from the Curie Weiss law, $\chi = C/(T-\theta_{CW})$, rearrangement (25), for $\theta_{CW} < 0$, yields $(C/|\theta_{CW}|)(1/\chi) -1 = (T/|\theta_{CW}|)$. Through scaling of T by $|\theta_{CW}|$, and the inverse susceptibility by C and $|\theta_{CW}|$ obtained from the fits, the resulting normalized plot, shown in Fig. 5 (omitting the -1 from the ordinate) allows direct comparison of the temperature dependent behavior of the two forms of $Mn_3Al_2Si_3O_{12}$, scaled for strength of near neighbor interactions and effective moments, also showing how they deviate from the Curie Weiss law relative to each other. The comparison shows that the crystalline phase deviates subtly from the Curie Weiss law starting at about $3|\theta_{CW}|$, showing lower susceptibility (larger scaled inverse susceptibility) than expected, suggesting the onset of antiferromagnetic fluctuations well above $T_N$. In contrast, the glass phase maintains Curie-Weiss behavior to substantially lower relative temperatures, until approximately $|\theta_{CW}|$, where its deviations are in the opposite sense than for the crystalline phase, showing higher susceptibility than expected from Curie-Weiss behavior, suggesting the presence of ferromagnetic fluctuations below $|\theta_{CW}|$. The temperature dependence of the fraction of ideal magnetic entropy loss on cooling (26), calculated by subtracting the measured entropy up to each temperature ($S_{Mn-Ca}$) from the total magnetic entropy observed in each case ($S_{Mn-Ca}$ at high temperature) and scaled by Rln6, is plotted vs. reduced temperature, $T/|\theta_{CW}|$, for both the crystal and glass forms of $Mn_3Al_2Si_3O_{12}$ is shown in the inset to Fig. 5. This scaled magnetic entropy plot indicates that the glass begins to lose a substantial fraction of magnetic entropy on cooling to $T = |\theta_{CW}|$ but that the crystal magnetic entropy is primarily lost at a lower relative temperature, at $T_N = |\theta_{CW}|/3$. However, as suggested by the scaled susceptibilities (main panel Fig. 5), the crystal also begins to lose entropy on the order $T = |\theta_{CW}|$, indicating that short range ordering on the lattice of corner sharing triangles in the crystalline garnet phase sets in near $T_N$, above the temperature of the long range ordering at $|\theta_{CW}|/3$.

**Conclusions**

We have characterized the differences between the magnetic behavior of crystalline and glass forms of a compound whose crystalline form is one of the well-known magnetically frustrating lattice geometries. For this specific case, where the crystal form is not strongly frustrated, the results show that the structural disorder in the glass yields a system exhibiting greater magnetic frustration. For systems where the crystal form is very frustrated, as in Kagome-based ferrites such as $SrGa_4Cr_8O_{19}$, where $\theta_{CW}/T_n > 200$ (1), the possible outcome of a similar comparison is not so obvious. For those cases, the challenge lies in the synthesis of a structural glass at the same compositions as the crystalline phases. Many of the systems whose geometric frustration is widely studied, e.g. the Kagome-based ferrites also include glass-like disorder of the magnetic connectivity due to partial replacement of magnetic ions in the frustrated lattice with non-magnetic equivalents (1). More detailed experimental study of the impact of that type of disorder in chemically and structurally well defined geometrically frustrated magnets would be of considerable future interest.

**Acknowledgements**

The research at Princeton was supported by the solid state chemistry program of the NSF, grant DMR-0703095. The work at Los Alamos was performed under the auspices of the US DOE. TMM gratefully acknowledges support from the National Science Foundation Graduate Research Fellowship Program.

**Figure Captions**

**Fig. 1.** (color on line) Main Panel: Comparison of the inverse magnetic susceptibilities, H/M, at an applied field of 1 T, for the crystalline and glass forms of $Mn_3Al_2Si_3O_{12}$. Inset, susceptibilities in the low temperature region showing the signature of the antiferromagnetic ordering in the crystal form.

**Fig. 2.** (color on line) Main Panel: The specific heats (per mole Mn or Ca) over a broad range of temperature for crystal and glass forms of $Mn_3Al_2Si_3O_{12}$ and $Ca_3Al_2Si_3O_{12}$. Inset – the temperature dependence of the difference in specific heats between the Mn and Ca versions of the garnet glasses, a measure of the contribution of the magnetic system. Also shown is the temperature dependence of the integrated magnetic entropy, which is somewhat less than Rln6, the amount expected if all magnetic entropy is lost in the glass on cooling down to 2 K.

**Fig. 3.** (color on line) Main panel: the temperature dependence of the integrated entropy for $Mn_3Al_2Si_3O_{12}$ in its crystalline garnet form, after the specific heats of the Ca analog grossular has been subtracted. The main panel and inset show the modeling of the excess specific heat at high temperatures in terms of the shift of one Einstein phonon mode to lower temperatures in the Mn analog. Comparison with experimental data is shown.

**Fig.4.** (color on line) Main Panel: the magnetic field dependence of the specific heat for the crystalline garnet form of $Mn_3Al_2Si_3O_{12}$ in the vicinity of the magnetic ordering transition. Inset: The same data for the structural glass.

**Fig. 5.** (color on line) Main panel: scaled inverse susceptibilities for crystal and glass forms of $Mn_3Al_2Si_3O_{12}$ allowing direct comparison of the magnetic properties. $\chi = C/(T-\theta_{CW})$, rearrangement yields $(C/|\theta_{CW}|)(1/\chi) - 1 = (T/|\theta_{CW}|)$, with C and $\theta_{CW}$ obtained from the high temperature fits (25). The solid line shows the ideal Curie-Weiss behavior, extrapolated from the

higher temperature data. Inset: The temperature dependence of the fraction of ideal magnetic entropy loss on cooling (26), plotted vs. $T/|\theta_{CW}|$, for both the crystal and glass forms of $Mn_3Al_2Si_3O_{12}$.

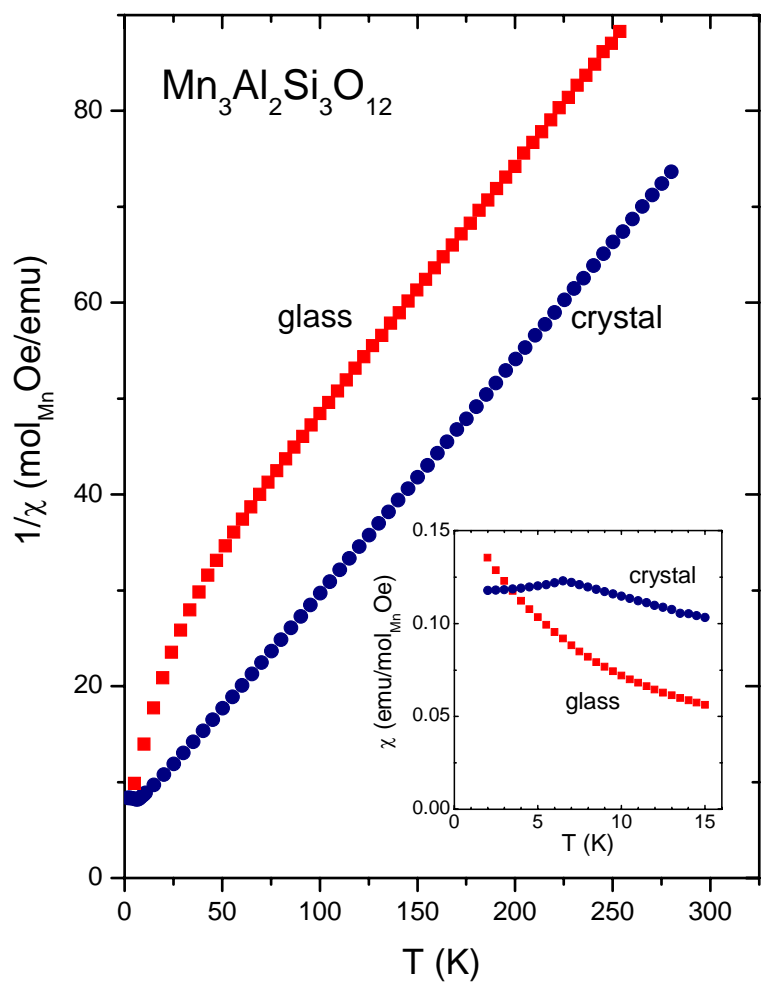

**Figure 1**

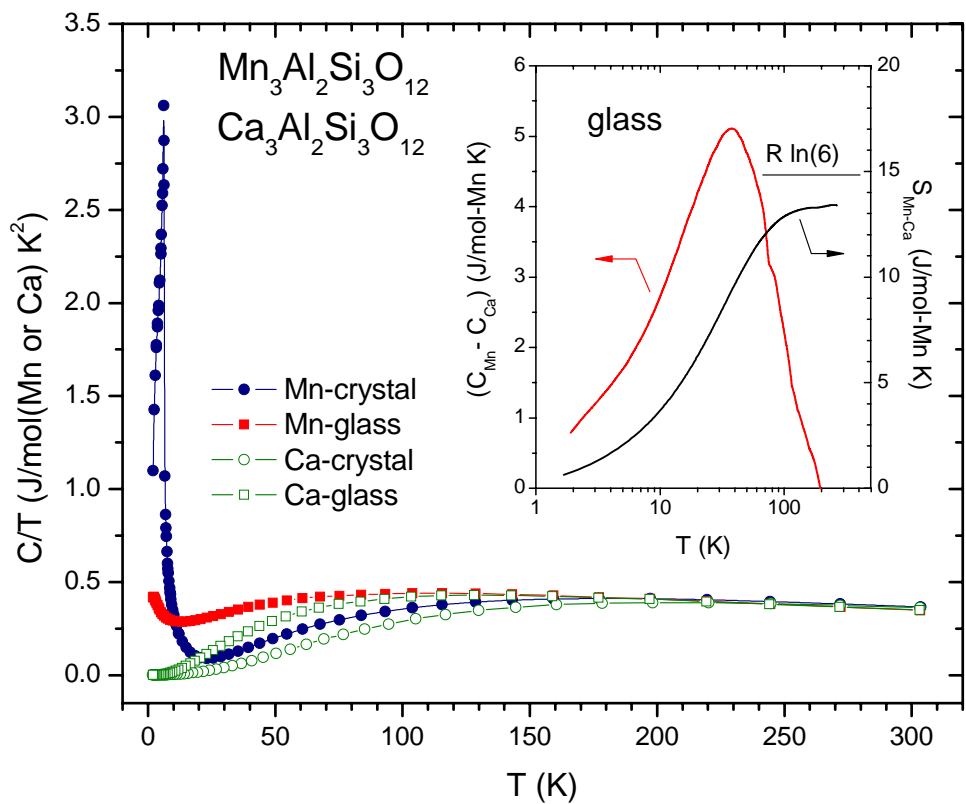

**Figure 2**

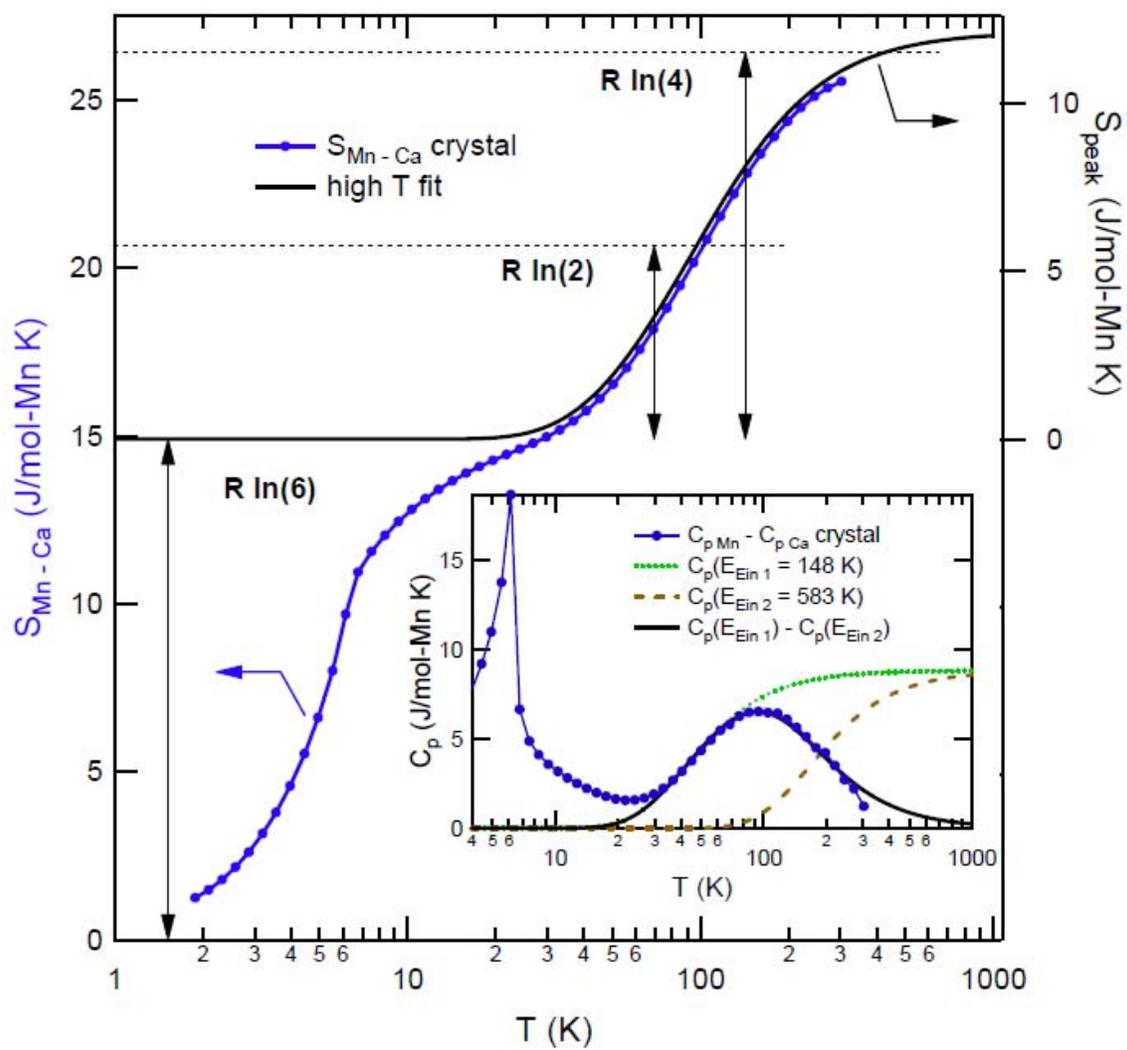

**Figure 3**

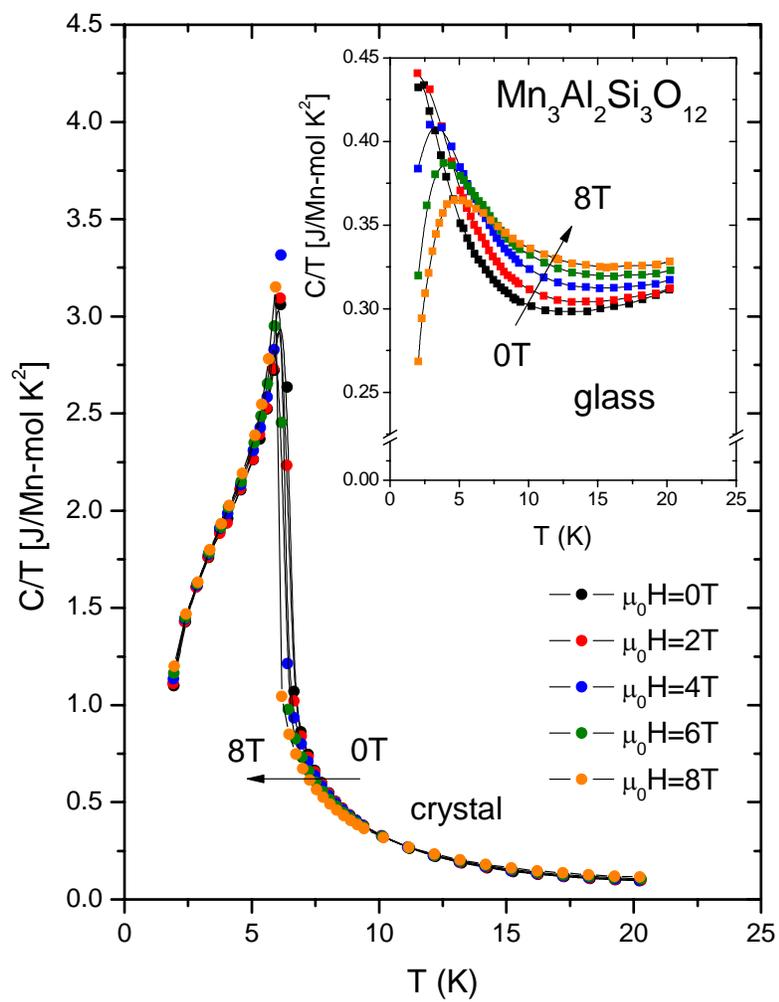

**Fig. 4**

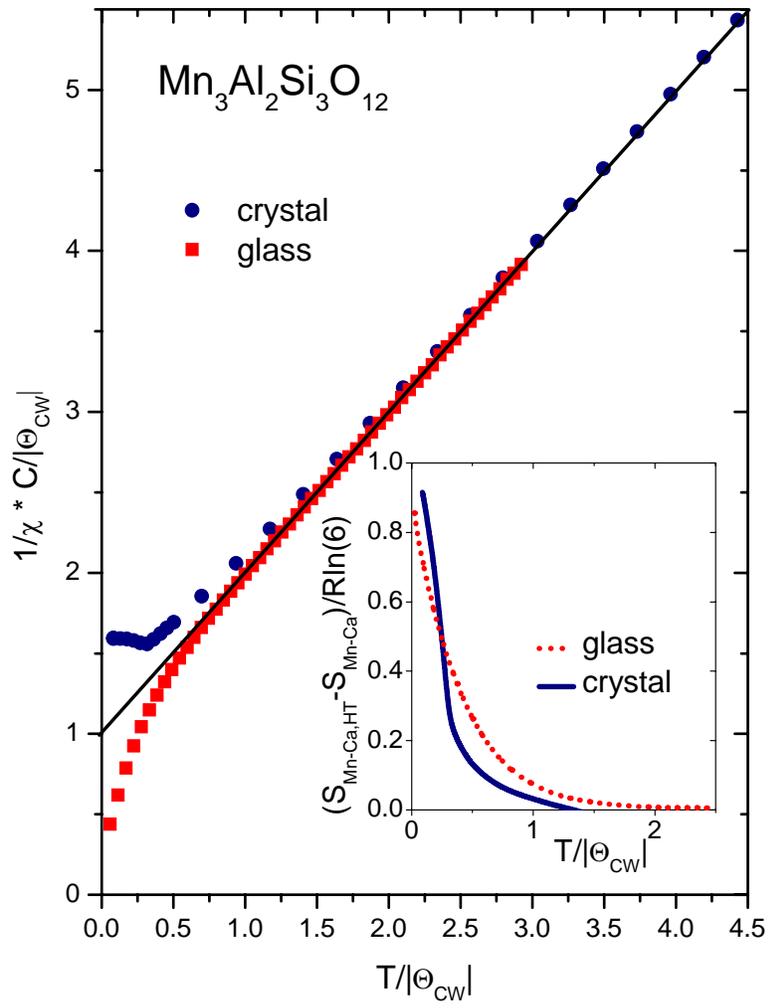

**Fig. 5**